\documentstyle[12pt]{article}
\begin{document}

\newcommand{\la}[1]{\label{#1}}
\newcommand{\re}[1]{ (\ref{#1})}
\newcommand{\rf}[1]{ Fig. \ref{#1}}
\newcommand{\nn}{\nonumber}
\newcommand{\ed}{\end{document}}
\newcommand{\be}{\begin{equation}}
\newcommand{\ee}{\end{equation}}
\newcommand{\ba}{\begin{eqnarray}}
\newcommand{\ea}{\end{eqnarray}}
\newcommand{\baz}{\begin{eqnarray*}}
\newcommand{\eaz}{\end{eqnarray*}}
\newcommand{\bb}{\begin{thebibliography}}
\newcommand{\eb}{\end{thebibliography}}
\newcommand{\ct}[1]{${\cite{#1}}$}
\newcommand{\ctt}[2]{$^{\cite{#1}-\cite{#2}}$}
\newcommand{\bi}[1]{\bibitem{#1}}

\textheight=22.0cm
\hsize=15.8 cm
\topmargin=-1.05 cm
\oddsidemargin=0.5cm

\begin{titlepage}
\vskip 3cm
\begin{center}
{\Large\bf Vacuum QCD and new information\\
 on nucleon
structure functions} \\[1cm]
{N.I.Kochelev}\\[1cm]
{Institut f\"ur Theoretische Physik, Freie Universit\"at Berlin, \\
 Berlin, Germany
\footnote{On leave of absence from Joint Institute for Nuclear Research,\\
Laboratory of High Energy,
Head Post Office P.O.Box 79, SU-101000 Moscow, Russia\\
E-mail: kochelev@dec1.physik.fu-berlin.de }}
\end{center}
\begin{center}
{\bf Abstract}\\[0.2cm]
\end{center}

It is shown, that specific chirality and flavor properties of the
 instanton-induced interaction allow us to explain the observed violation
Ellis-Jaffe and Gottfried sum rules.
A large value of these violations originates from
the anomalous growth of the instanton-induced interaction between
quarks at high energy.  The new  sum rule which connect the
values of violations of Ellis-Jaffe and Gottfried
 sum rules have been obtained.
\end{titlepage}

At present, the possibility of anomalous baryon-number violation in
electroweak interaction at high energy is widely under discussion (see
\ct{a1}).
It was known that t'Hooft interaction \ct{a2} can be treated as the
mechanism for baryon-number violation. This interaction is induced by
vacuum fluctuations of gauge fields-instantons. However, the
t'Hooft's instanton amplitude  includes a very small semiclassical
factor $ exp[-2\pi/\alpha_W] $ and one would think that there are
no real possibilities
 for observing this violation experimentally.
The renewal of interest in this problem happens basically due to the fact
that at high energy the multiple production of gauge bosons from
instanton leads to
anomalous growth of the baryon-number violation cross-section \ct{a3}.

In QCD, the instantons violate the chirality and therefore the same
mechanism, evidently leads to anomalous quarks chirality violation
at high energy.
The most interesting fact is that in contrast of problems of the experimental
 searches of baryon-number violation, in strong interaction we  already
 observe multiple anomalous spin effects, which can be connected only with
 a large quarks chirality violation at high energy (see \ct{a4}).

The  perturbative QCD predicts the disappearance of chirality violation
at high energy and therefore in \ct{a5}
\footnote{In papers \ct{a6} the instanton mechanism for
chirality violation in QCD was also discussed.}the instanton interaction
have been proposed as the fundamental mechanism for the spin effects at
high energy.

The very interesting results on polarized \ct{a7}, \ct{a8} and unpolarized
 \ct{a9} nucleon structure functions have been published recently.
 These experimental results lead to the conclusion that the behaviour of the
 sea quarks at the small $ x$   differs strongly from the perturbative QCD
 predictions. Just the data analysis shows that the quark sea is strongly
 polarized and it is asymmetric in flavor. The very fast growth of quark
 sea at low $x$ is still a very mysterious fact.

Here, we will present the arguments that these surprising results can be
explained by manifestation of a complicated structure of QCD vacuum,
namely by taking into account the instanton-induced interaction between
quarks.

Indeed, it is well known \ct{a10}, that behavior of quark-quark cross
section at high energy determines the behavior of the nucleon structure
functions at low $x$. So it immediately follows that if the quark-quark
cross section at high energy have the behaviour $\sigma\sim s^{\alpha-1}$,
 then the distribution functions of sea quarks is $ q(x)\sim 1/x^\alpha$
at low $x$ . Thus the anomalous behaviour of the quark-nucleon
 cross-sections should provide the anomalous behaviour of the structure
functions at $x\rightarrow0$. Unfortunately, the exact dependence of the
instanton-induced interaction from energy is still unknown \ct{a11}.
It is connected with the fact that at very high energy
the small size instanton approximation is incorrect and then
one needs to take into account
the overlapping of instantons.

Here we suppose that QCD instantons induce the following form for growth
of cross section
\footnote{This form of the cross section behaviour rise from the estimates
in the small size instanton approximation \ct{a11}.}
\be
\sigma_{qq}\sim s^{\alpha_I-1},
\la{e1}
\ee
where $\alpha_I>1 $ and we will find the value for $\alpha_I$ from
 the fit of the experimental data
on the structure functions.

The peculiarity of the t'Hooft lagrangian \ct{a2}
\footnote{For simplicity, we present here only the instanton-induced
 lagrangian for case $N_f=2$. The case of $N_f=3$ was considered in ref.
 \ct{a12}.}
 \begin{eqnarray}
{\cal L}_{eff}^{(2)}(x)= \int d\rho\thinspace
 n(\rho)(\frac{4}{3}\pi^2\rho^3)^2\left\{\right.\relax
\bar u_Ru_L\bar d_Rd_L\left.\right. \relax
[1+\frac{3}{32}(1-\frac{3}{4}\sigma_{\mu\nu}^u\sigma_{\mu\nu}^d) \relax
\lambda_u^a\lambda_d^a]+(R\longleftrightarrow L)\left.\right\},
\la{e4}
\end{eqnarray}
 where $\rho$ is the size of instanton and $n(\rho)$ is its density, consists
 in the specific chirality and flavor properties of the interaction
 \re{e4}. Namely, it changes chirality by the value $\Delta Q=-2N_f$ and
 it acts only between different flavors of the quarks. In fact, these
 properties are the manifestation of the Pauli exclusion principle for
  the scattering
 of quarks off fermion zero modes in the instanton field.

One of the diagrams coming from the instanton-induced interaction
to the nucleon structure functions at $x\rightarrow0$ is shown in Fig.1.
One can obtain  more complicated diagrams  from  Fig.1
using the insertion of the additional instanton-antiinstanton pairs.

It was shown in \ct{a5} that the contribution from diagram Fig.1
to the structure functions violates both the Ellis-Jaffe sum rule \ct{a13},
related to the value of chirality carried by proton quarks, and the
Gottfried sum rule \ct{a14} connected with flavor properties of the quark sea.
However, not enough fast growing instanton-induced asymptotic of structure
functions at low $x$ ($q(x)\sim1/xln^2x$) was used in \ct{a5}.
Due to this fact, this asymptotic does not describe the observed growth
of the quark sea at low $x$.

Here, taking into account \re{e1}, we present the new
 form of the asymptotic
\be
q(x)_{low{\ } x}\sim\frac{1}{{x^{\alpha_I}}}
\la{e5}
\ee
where $\alpha_I> 1$. From unitarity  constraints \ct{a15} which lead to
 the
instanton-induced cross-section  $\sigma_{qq}\sim exp(-E/E_0)$  \ct{a11}
at very high energy,

 we get also
\be
q(x)_{x\rightarrow 0}\rightarrow exp(-x_0/x)
\la{e6}
\ee
at very low $x$.

 In \ct{a5} it was shown that instantons lead
 to the hard
sea quarks distribution functions, similar to valence form at $x\rightarrow1$.
This hard distribution is caused by the smaller average size of instantons in
 QCD vacuum as compared with the confinement radius ($\rho\approx R/3$).
 Due to the same reason, the instanton-induced quark sea is additive,
 and   its distribution function is proportional to the
  distribution function of the valence quark from which it was produced.

   For the quark-quark scattering due to  instantons,
 the chirality and flavor properties of the sea quarks are determined by
 the valence quark lines incoming t'Hooft vertex (Fig.1) \ct{a5}.
 So, to obtain the relation between the various chirality and flavor sea
  quark distribution functions it is enough to consider the contribution
  to the quark sea from diagram of Fig.1.
 Let us consider the quark sea created by interaction \re{e4} from the
 proton $SU(6)_W$ symmetrical wave function:
 \be
 p\uparrow=\frac{5}{3}u\uparrow+\frac{1}{3}u\downarrow+
 \frac{1}{3}d\uparrow+\frac{2}{3}d\downarrow.\nn
 \ee
We will use the properties of the vertex \re{e4} which lead to
the opposite chirality and different flavor as compare with the chirality and
 flavor of the initial valence quark. In this way one can
  easily
 obtain the following expression for the chirality and
 flavor sea quark distribution
 functions  $q_{+(-)}$:
 \ba
 2\bar u^I_+(x)&=&\frac{2}{3}\frac{N_Id_v(x)x^{\alpha_v}}{x^{\alpha_I}}f_u(x)
{\ },
 {\ }{\ }{\ }\hspace*{0.5cm}
 2\bar
u^I_-(x)=\frac{1}{3}\frac{N_Id_v(x)x^{\alpha_v}}{x^{\alpha_I}}f_u(x),\nn\\
 2\bar d^I_+(x)&=&\frac{1}{6}\frac{N_Iu_v(x)x^{\alpha_v}}{x^{\alpha_I}}f_u(x)
{\ },
 \hspace*{0.5cm}{\ }{\ }{\ }
 2\bar
d^I_-(x)=\frac{5}{6}\frac{N_Iu_v(x)x^{\alpha_v}}{x^{\alpha_I}}f_u(x),\nn\\
2\bar s_+^I(x)&=&2(\bar u_+^I(x)+\bar d_+^I(x)),\hspace*{1.3cm}2\bar
s_-^I(x)=2(\bar u_-^I(x)+\bar d_-^I(x)),
\la{e7}
\ea
where $N_I$  is any constant,$ {\ }u_v(x),{\ }d_v(x)$ are the valence
 distribution functions,
which we choose in the usual form :
 \be
u_v(x)=N_ux^{-\alpha_v}(1-x)^{bu},\hspace*{0.2cm}d_v(x)=
N_dx^{-\alpha_v}(1-x)^{bd}.
\la{e8}
\ee
In the formula \re{e7} the function $f_u(x)$ provides the correct unitarity
 limit
at $x\rightarrow 0$ . We can choose this function in
the form
\ba
f_u(x)=\cases{ 1,&if $x> x_0$;\cr
 exp(-(x_0/x-1)),&$x < x_0$.\cr}
\label{e9}
\ea

Instantons also lead to the valence quark chirality fliping due to the
annihilation diagram which is shown in Fig.2. Its contribution to
the valence quark distributions functions is:
 \ba
  u^I_{v+}(x)&=&\frac{5}{6}\frac{N_Iu_v(x)x^{\alpha_v}}{x^{\alpha_I}}f_u(x){\
},
 {\ }{\ }{\ }\hspace*{0.5cm}
  u^I_{v-}(x)=\frac{1}{6}\frac{N_Iu_v(x)x^{\alpha_v}}{x^{\alpha_I}}f_u(x),\nn\\
  d^I_{v+}(x)&=&\frac{2}{3}\frac{N_Id_v(x)x^{\alpha_v}}{x^{\alpha_I}}f_u(x){\
},
 \hspace*{0.5cm}{\ }{\ }{\ }
 d^I_{v-}(x)=\frac{1}{3}\frac{N_Id_v(x)x^{\alpha_v}}{x^{\alpha_I}}f_u(x),
\la{e10}
\ea
The quark sea induced by the perturbative gluons does not differ in chirality
or  flavor, and therefore  it can be taken in the  form using
quark-counting rules:
\be
\bar u(x)_{+,-}^p=\bar d(x)_{+,-}^p=2\bar s(x)_{+,-}^p=
N_s(1-x)^5/x^{\alpha_P(0)-1},
\la{e11}
\ee
where the suppression of the strange sea one usually connects with the
belated start of the evolution due to the large mass of the
 strange quark \ct{a16} and $\alpha_p(0)=1.08$ is the intercept of the
soft pomeron \ct{a17}.

{}From \re{e7}-\re{e11} one can obtain the unpolarized
$\bar q(x)=\bar q_+(x)+\bar q_-(x) $  and the polarized
$\Delta\bar q(x)=\bar q_+(x)-\bar q_-(x) $
structure functions. So, the nucleon structure functions $ F_2^{\mu N}(x)$
have the
following form:
\ba
F_2^{\mu p}(x)=e_u^2( u_v^r(x)+2\bar u^p(x))+
e_d^2 (d_v^r(x)+2\bar d^p(x))+2e_s^2\bar s^p(x)+(\sum_q e_q^2)A(x),\nn\\
F_2^{\mu n}(x)=e_d^2( u_v^r(x)+2\bar u^p(x))+
e_u^2 (d_v^r(x)+2\bar d^p(x))+2e_s^2\bar s^p(x)+(\sum_q e_q^2)A(x),
\la{e12}
\ea
where
\be
A(x)=\frac{N_I(u_v(x)+d_v(x))x^{\alpha_v+1}}{x^{\alpha_I}}f_u(x),
\la{e13}
\ee
 and $q_v^r(x)$ is renormalized valence quark distribution
function . The renormalization  takes into account the instanton
contribution to valence quark distribution functions:
\ba
\int_0^1dx(u_v^r(x)+u_v^I(x))=2,\nn\\
\int_0^1dx(d_v^r(x)+d_v^I(x))=1.
\la{e14}
\ea
Fits of the  NMC and H1 experimental data \ct{a9} for the unpolarized structure
function of the proton $F_2^{p}(x)$ and the neutron-proton ratio
$F_2^{\mu n}(x)/F_2^{\mu p}(x)$ is shown in Fig.3, 4.

The values of the parameters obtained in our fit is
:
\be
bu=2.79;{\ } bd=3.46;{\ }\alpha_v=0.37;{\ }N_s=0.095;{\ }N_I=0.0046;{\ }
x_0=0.0005;{\ }\alpha_I=1.36 .\nn
\ee
Thus, we obtain a large excess of ${\ }\alpha_I$  a compared with 1,
 which provides the growth of $F_2^{N}(x)$
at low $x$.

The contribution of instanton-induced sea   to the total
sea is shown in Fig.5.
{}From this picture we  conclude,
that the nonpertubative sea gives the large contribution to quark sea
distribution functions
in the interval $ 0.0001<x<0.1$
 and the growth of the structure functions
  in this region is determined by the
anomalous dependence on the energy of the instanton-induced interaction
 between quarks.

The flavor dependence of the instanton sea
\re{e7}leads to large violation of the $ SU(2)$-flavour symmetry quark sea,
which, of course, provides a strong violation of the Gottfried sum rule
 for
the integrated difference of the structure functions
\be
I_G(t)=\int_t^1\frac{dx}{x}(F_2^{\mu p}(x)
-F_2^{\mu n}(x)),\quad I_G(0)=\frac{1}{3}.
\nn
\ee
Our model predicts a very large violation of the
 Gottfried sum rule: $I_G^I(0)=0.247$ which can be compared with
the NMC  result \ct{a9}:
$$I_G^{NMC}=0.257\pm0.017.$$

Further, since we have chirality distribution functions,
 we can obtain the spin-dependent structure functions.
It is only necessary to introduce chirality distribution functions for valence
quarks.
Let us choose these distributions in the form given by the Carlitz-Kaur model
\ct{a18}
\ba
\Delta u_v(x)&=&(u_v(x)-2d_v(x)/3)cos\Theta_D(x)\nn\\
\Delta d_v(x)&=&-d_v(x)cos\Theta_D(x)/3,
\la{e16}
\ea
where $ cos\Theta_D(x)$  takes into account the depolarization of
 the valence quarks in the low $ x $-region  and it
  can be interpreted as the contribution of confinement
   forces to the valence quark chirality violation.
It has the following form
\be
cos\Theta_D(x)=(1+H_1(1-x)^2/x^{\alpha_v})^{-1},
\la{e17}
\ee
which provides the correct asymptotic for the spin-dependent valence
 structure functions at $x\rightarrow 0$.
This asymptotic form is determined by contribution from
 $A_1$ trajectory with intercept $\alpha_{A_1}(0)\approx0$.
{}From the formulas \re{e7}-\re{e10}, and \re{e16}, \re{e17}
 one can obtain the
 proton and neutron structure functions $g_1^{p,n}(x)$:
\ba
g_1^p(x)=\frac{4}{18}\Delta u_v(x)+
\frac{1}{18}\Delta d_v(x)-(\sum_q e_q^2)B(x),\nn\\
g_1^n(x)=\frac{4}{18}\Delta d_v(x)+
\frac{1}{18}\Delta u_v(x)-(\sum_q e_q^2)B(x),
\la{e18}
\ea
where
\be
B(x)=\frac{(N_I(2u_v(x)-d_v(x))x^{\alpha_v+1}}{6x^{\alpha_I}}f_u(x),
\la{eee13}
\ee

In Fig.6,7 we present the results for $g_1^p(x)$ and for $g_1^n$ .
 The value of the parameter $H_1=0.14$   was determined from
 normalized to the Bjorken sum rule \ct{a19} .
In Fig.6 the contribution to $g_1^p(x)$
 from the valence quarks is also shown.
This contribution does not describe the data.
The first moment of $g_1^p(x)$  is,
\be
I_p=\int_0^1g_1^p(x)dx=-0.03,
\la{ee18}
\ee
which leads to very large violation of the Ellis-Jaffe sum rules \ct{a14}: $
I_p^{EJ}=0.175\pm0.018$.

The neutron spin-dependent structure function
$g_1^n(x)$ is  negative and it reaches large absolute values at low $ x $.
The first moment is
$I_n=-0.238, $
which is much smaller than the predictions of most models.
In Fig.8  we also show the result of our calculation of the deutron
 spin-dependent structure
functions $g_1^d(x)$.

{}From Fig.6-8 it follows that our model predicts a
 very fast decrease of the spin-dependent structure functions $g_1(x)$
  to  negative values of decreasing $x$.
It is connected with shadowing of the valence
 quark chirality by the sea quark chirality and also with fliping the valence
quark chirality
 by instantons.
 It should be stressed that
  in our model the growth of the unpolarized structure
   functions $F_2^N(x)$  in the low $ x $ regions is directly connected
    with the decrease of the spin-dependent structure function $g_1^p(x)$.

For values of chirality, induced by instantons,
 one obtains from \re{e7}, \re{e10}:
\be
\Delta u_I=0.29 {\ }
;{\ }\Delta d_I=-0.96 {\ };{\ }
\Delta s_I=-0.72 .
\nn
\ee
The valence proton quark chiralities are $
\Delta u_v=1.01{\ };{\ }\Delta d_v=-0.24 .$
Therefore,  the total chirality, carried proton quarks, is:
\be
\Sigma_{tot}^I=\Delta u+\Delta d+\Delta s=-1.4 .
\la{e21}
\ee
This value  contradicts the EMC result \ct{a7}:
\be
\Sigma^{EMC}=0.12\pm0.24 .
\la{e22}
\ee
The cause for this discrepancy is the different
extrapolation to low $x$. So, the quark chirality for the EMC interval is:
$\Sigma(0.01<x<0.7)=0.24 {\ }.$
 It means that the main contribution in value \re{e21}
due to the kinematical region which is not accessible in the EMC experiment.

{}From \re{e7}, {\re{e10} at $SU(6)_W$ limit for the valence quarks structure
functions
$u_v(x)=2d_v(x)$ we obtain a very interesting relation between of the
values of the violation Gottfried and Ellis-Jaffe  sum rules:
\be
\int_0^1\frac{dx}{x}(F_2^{\mu p}(x)-F_2^{\mu
n}(x))-\frac{1}{3}=\int_0^1g_1^p(x)dx-\frac{5g_A^8+3g_A^3}{36}.\nn
\ee

Thus we suggest a new mechanism for the explanation of the anomalous
behaviour of the polarized and unpolarized structure functions al low $ x$.
It is related to the growth of the  instanton-induced quark-quark
interaction at high energy.
So, the  instanton exchange leads to the fast growth in the unpolarized
$F_2^N(x)$ and to the fast decrease of the polarized
$g_1^N(x)$ functions at $0.0001<x<0.1$.

We have also shown, that for the  exchange by instantons
the Pauli principle for quarks on zero fermion modes provides
 strong correlation of spin and flavor of the sea quarks and valence quarks.
This correlation leads to violation of the Ellis-Jaffe and Gottfried
  sum rules.
The large value of these violations is determined by the large
contribution of the  instanton-induced sea in the low $0.0001<x<0.1$ region.

The author are sincerely thankful to A.E.Dorokhov, A.V.Efremov, N.N.Nikolaev,
G.Ramsey, A.Sch\"afer, E.V.Shuryak  for useful discussions,
 Prof.Meng Ta-chung for warm hospitality at
 the Free University of Berlin   and the DFG (Project: ME 470/7-1)
  for financial support.

\bb{99}
\bi{a1} M.P.Mattis {\it Phys.\ Rep.\ } {\bf 214} (1992) 156.
\bi{a2} 't Hooft {\it Phys. Rev.} {\bf D14} (1976) 3432.
\bi{a3} A.Ringwald {\it Nucl.\ Phys.\ } {\bf B330} (1990) 1;\\
O.Espinosa {\it Nucl.\ Phys. } {\bf B343} (1990) 310.
\bi{a4}G Bunce et al. {\it Part.\ World\ }{\bf 3} (1992) 1 .
\bi{a5} A.E.Dorokhov,N.I.Kochelev
 {\it Phys.\ Lett.\ } {\bf B259} (1991) 335;\\
  {\it Phys.\  Lett.\ } {\bf B304} (1993) 167;
  {\it Int.\ J.\ of\ Mod.\ Phys.\ } {\bf A8} (1993) 603.
\bi{a6} S.Forte, {\it Phys.\ Lett.\ }{\bf B224} (1989) 189;
{\it Nucl.\ Phys.\ }{\bf B331} (1990) 1;\\
K.Steininger, W.Weise {\it  Phys.\ Rev.\ }{\bf D48} (1993) 1433.
\bi{a7} EMC, J.Ashman et al.  {\it Nucl.\ Phys.\ }{\bf B328} (1990) 1.
\bi{a8} SMC B.Adeva et al. {\it Phys.\ Lett.\ } {\bf B302} (1993) 533;\\
 E-142 Collaboration {\it Phys.\ Rev.\ Lett.\ }{\bf 71} (1993) 959.
\bi{a9} NMC, P.Amaudruz et al.
 {\it Phys.\ Rev.\  Lett.\ }{\bf 66} (1991) 2712;\\
 Preprint CERN-PPE/93-1993; H1 Coll. Preprint DESY 93-113.
\bi{a10} P.V.Landshoff, J.C.Polkinghorne, R.D. Short {\it Nucl.\ Phys.\ }
{\bf B28} (1971)~225.
\bi{a11} I.I Balitsky,M.G.Ryskin {\it Phys.\ Lett. } {\bf B296} (1992) 185.\\
The contribution of the small size instantons to the coefficient function
in front of parton distributions for DIS was calculated in paper
I.I.Balitsky, V.M.Braun {\it Phys.\ Rev.\ }{\bf D47} (1993) 1879.
\bi{a12} M.A.Shifman, A.I.Vainstein, V.I.Zakharov {\it Nucl.\ Phys.\ }
{\bf B163} (1980)~46.
\bi{a13} J.Ellis, R.L.Jaffe, {\it Phys.\ Rev.\ }{\bf D9} (1974) 1444.
\bi{a14} K.Gottfried {\it Phys.\ Rev.\ Lett.\ } {\bf 18} (1967) 1154
\bi{a15} V.I.Zakharov {\it Nucl.\ Phys.} {\bf B353} (1991) 683.
\bi{a16}V.Barone at el. {\it Phys.\ Lett.\ } {\bf B268} (1991)279.
\bi{a17} A.Donnachie, P.V.Landshoff {\it Nucl.\ Phys.\ }{\bf B231} (1984)~189.
\bi{a18} R.D.Carlitz, J.Kaur {\it Phys.\ Rev.\ Lett.\ }{\bf 38} (1977) 673.
\bi{a19} J.D.Bjorken, {\it Phys.\ Rev.\ } {\bf 148} (1966) 1467.

\eb
\end{document}